\documentclass[conference]{IEEEtran}
\IEEEoverridecommandlockouts
\usepackage{cite}
\usepackage{amsmath,amssymb,amsfonts}
\usepackage{algorithmic}
\usepackage{graphicx}
\usepackage{subfig}
\usepackage{textcomp}
\usepackage{xcolor}
\def\BibTeX{{\rm B\kern-.05em{\sc i\kern-.025em b}\kern-.08em
    T\kern-.1667em\lower.7ex\hbox{E}\kern-.125emX}}

  \makeatletter
    \newcommand{\linebreakand}{%
      \end{@IEEEauthorhalign}
      \hfill\mbox{}\par
      \mbox{}\hfill\begin{@IEEEauthorhalign}
    }
    \makeatother

\begin{document}

\title{Attacks on Continuous Chaos Communication and Remedies for Resource Limited Devices
}

\author{
\IEEEauthorblockN{Rahul Vishwakarma}
\IEEEauthorblockA{\textit{Computer Eng. \& Computer Sci.} \\
\textit{California State Uni. Long Beach}\\
Long Beach, CA, USA \\
rahuldeo.vishwakarma01@student.csulb.edu}
\and

\IEEEauthorblockN{Ravi Monani}
\IEEEauthorblockA{\textit{Electrical Engineering} \\
\textit{California State Uni. Long Beach}\\
Long Beach, CA, USA \\
ravi.monani01@student.csulb.edu} 
\and

\IEEEauthorblockN{Amin Rezaei}
\IEEEauthorblockA{\textit{Computer Eng. \& Computer Sci.} \\
\textit{California State Uni. Long Beach}\\
Long Beach, CA, USA \\
amin.rezaei@csulb.edu} 
\linebreakand 

\IEEEauthorblockN{Hossein Sayadi}
\IEEEauthorblockA{\textit{Computer Eng. \& Computer Sci.} \\
\textit{California State Uni. Long Beach}\\
Long Beach, CA, USA \\
hossein.sayadi@csulb.edu}
\and

\IEEEauthorblockN{Mehrdad Aliasgari}
\IEEEauthorblockA{\textit{Computer Eng. \& Computer Sci.} \\
\textit{California State Uni. Long Beach}\\
Long Beach, CA, USA \\
mehdad.aliasgari@csulb.edu}
\and

\IEEEauthorblockN{Ava Hedayatipour}
\IEEEauthorblockA{\textit{Electrical Engineering} \\
\textit{California State Uni. Long Beach}\\
Long Beach, CA, USA \\
ava.hedayatipour@csulb.edu}
}

\maketitle

\begin{abstract}
The Global Wearable market is anticipated to rise at a considerable rate in the
next coming years and communication is a fundamental block in any wearable device. In communication, encryption methods are being used with the 
aid of microcontrollers or software implementations, which are power-consuming and incorporate complex hardware implementation. 
Internet of Things (IoT) devices are considered as resource-constrained devices that 
are expected to 
operate 
with low computational power and resource utilization criteria. 
At the same time, recent research has shown that IoT devices are highly vulnerable to emerging security threats, which elevates the need for low-power and small-size hardware-based security countermeasures. Chaotic encryption is a method of data encryption that utilizes chaotic systems and non-linear dynamics to generate secure encryption keys. It aims to provide high-level security by creating encryption keys that are sensitive to initial conditions and difficult to predict, making it challenging for unauthorized parties to intercept and decode encrypted data. Since the discovery of chaotic equations, there have been various encryption applications associated with them. 
In this paper, we comprehensively analyze the physical and encryption 
attacks on continuous chaotic systems in resource-constrained devices and their potential remedies. 
To this aim, we introduce different categories of attacks of chaotic encryption. 
Our experiments focus on chaotic equations implemented using Chua's equation and 
leverages circuit architectures and provide simulations proof of remedies for different attacks. 
These remedies are provided to block the attackers from stealing users' information (e.g., a pulse message) 
with negligible cost to the power and area of the design.

\end{abstract}


\section{Introduction}
With advancements in microelectronic circuit technology and the miniaturization of sensors, the possibility of developing smaller and more reliable Implantable/Wearable Devices (IWDs) is progressive. The Integrated Circuit (IC) design companies have broadened their portfolios to be well composed of these IWDs. These devices are designed to accomplish designated tasks such as monitoring the oxygen level in the blood (Pulse Oximetry test), the glucose level in the blood, the heartbeat, body temperature, hydration level, etc. The impressive improvement in reliability and efficiency of IWDs such as smartwatches, fitness bands, and body patches has brought the medical industry and the IWD manufacturers together to obtain approval as a medical device from the US Food and Drug Administration (FDA). \cite{medcity} The IWDs provide the real-time data accumulated from the on-chip sensors and enable the utilization of those. These data are collected remotely and can be managed through smartphones. The health organization accesses the data remotely to monitor the patients and keep that process cost-effective. \cite{pwc} Thus, the accessibility of the features and health data provided by these IWDs has made them functional for many people. \cite{ErdmierCasey2016Wdii}

Wearable Technology Market (WTM) is expected to grow to USD 265.4 billion in 2026 from USD 116.2 billion in 2021 at a rate of 18.0 \% CAGR (Compound Annual Growth Rate)1 \cite{market}. Wearable devices develop a large consumer base when they are equipped with capabilities such as adaptability, reliability, portability, proper form factor, fault resilience, etc. The devices with these capabilities in the industry top the list of smart wearable devices \cite{MudraBand, jbuds, Bioheart, Iot}. These devices generate a variety of vast and potentially useful data that can be collected and analyzed by wearable device manufacturers \cite{ErdmierCasey2016Wdii}. Due to the fact that these devices function with wireless networks like NFC, Zigbee, and Bluetooth, they are vulnerable to cyberattacks.

The informative data collected from devices cannot be classified as secure without certain security regulations or encryption algorithms for the welfare of the consumer. Synchronization of chaos is the foundation for utilizing chaos in communications. The raw and unencrypted data is transformed into a scrambled signal by the chaotic transmitter prior to being transmitted via a public channel, which can either be wireless, like in the case of sensor networks in wearable devices or IoT, or wired, like in power grids. The traditional encryption algorithms for communication are symmetric and asymmetric encryption. In symmetric encryption, just one key is involved, a private key that needs to be sent separately with the highest level of security. In the latter case, a pair of encryption keys are involved, and the receiver has a private key to decrypt the message.

The remainder of this paper is organized as follows. First, Chua's model is introduced in Section II. Some of the known security attacks that IWDs can face are analyzed and explained in Sections III and IV. The experimental and simulation results follow the attack explanations in Section III for physical attacks and Section IV for encryption attacks, and the paper concludes with Section V.

\begin{table}[!t]
	\centering
	\caption{Cryptographic Algorithms}
	\vspace*{0.02in}
	\begin{tabular}{ c | c }
		\hline
		Algorithm & Purpose \\ 
		\hline
            \hline
		Advanced Encryption Standard (AES) & Confidentiality \\
            \hline
	    Rivest-Shamir-Adelman (RSA) and & Digital Signatures   \\
     Elliptic Curve Cryptography (ECC) & Key Transport  \\
     \hline
		Diffie-Hellman (DH) & Key Agreement\\
  \hline
		SHA-1 and SHA-256 & Integrality\\
		\hline
	\end{tabular}
	\label{tab:algo}
\end{table}

\section{Encryption Using Chua's Circuit}
The discovery of chaos is considered a major breakthrough when it comes to ciphering. There have been multiple attempts to use chaos in the fields of encryption \cite{daniel,esteban, wei, vaidya, 9420744, 8617960}. However, the objective still remains challenging when we aim to bring chaos to the CMOS level. In \cite{hedayatipour2021comprehensive} different modes of chaotic equations have been compared for various performance metrics, and in \cite{9955334} an encryption architecture suitable for communication of on-chip sensors using different chaotic equations is implemented.

The algorithms listed in Table \ref{tab:algo} can be used depending on the target Internet of Things (IoT) application, as they vary in resources such as memory and power. The symmetric encryption administers the shared private key between sender and receiver, whereas in the asymmetric encryption, the sender provides the public information and the receiver decrypts that with the private information \cite{646128}. The usually used symmetric encryption is the Advanced Encryption Standard (AES) block cipher \cite{alma991012566379702910}. However, AES is not considered a guaranteed reliable communication \cite{1190590}. Asymmetric encryption was introduced by Diffie and Hellman and is also called public-key cryptography \cite{alma991072661417602901}. Famous asymmetric encryptions are Rivest-Shamir-Adelman (RSA) and the Diffie-Hellman (DH) asymmetric key agreement algorithm \cite{6188257}. The downside of these algorithms is that they require more hardware utilization due to their computational requirements and factorization complexity \cite{8079618}.

There are determined researches claiming that the traditional encryption can be broken with way less textit{qubits} compared to the existing encryption \cite{proos2003shor}. The thorough analysis of post-quantum era and the block-chain cryptography to quantum computing attacks has been done in literature \cite{8967098}. There are no post-quantum block-chain algorithms available at present that can provide low-power implementation, low computational complexity, and lower execution time. These are critical characteristics for any resource-constrained devices like IWDs \cite{8932459}. Therefore, methods of encryption other than symmetric and asymmetric security are gaining in importance and quickly becoming necessary. Starting to implement chaos on the hardware, this paper focuses on implementing Chua's circuit and its robustness against various attacks. 

 \begin{figure}[!t]
 \centering 
\subfloat[]{\includegraphics[width=0.85\columnwidth]{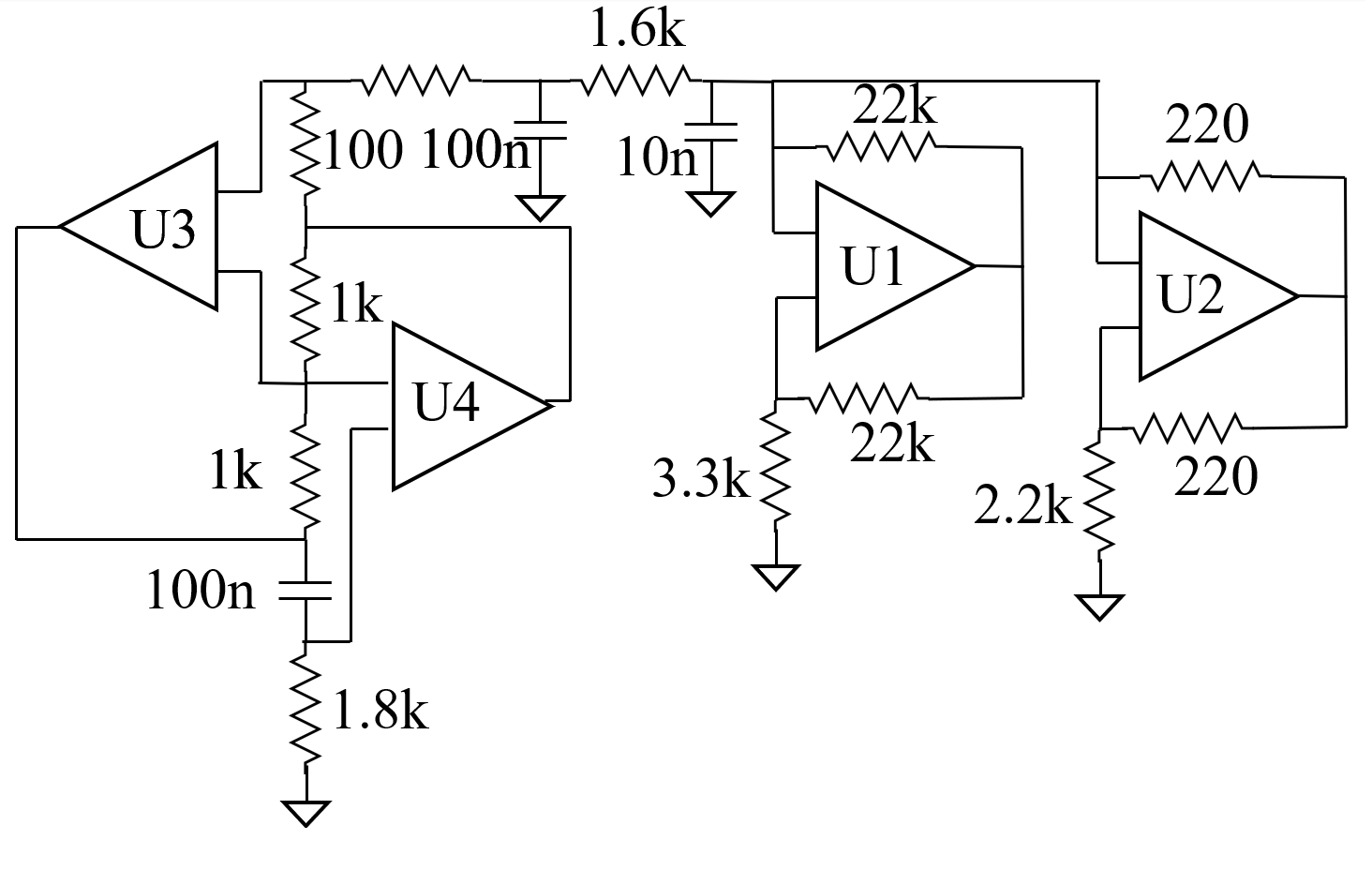}} \\ 
\subfloat[]{\includegraphics[width=0.75\columnwidth]{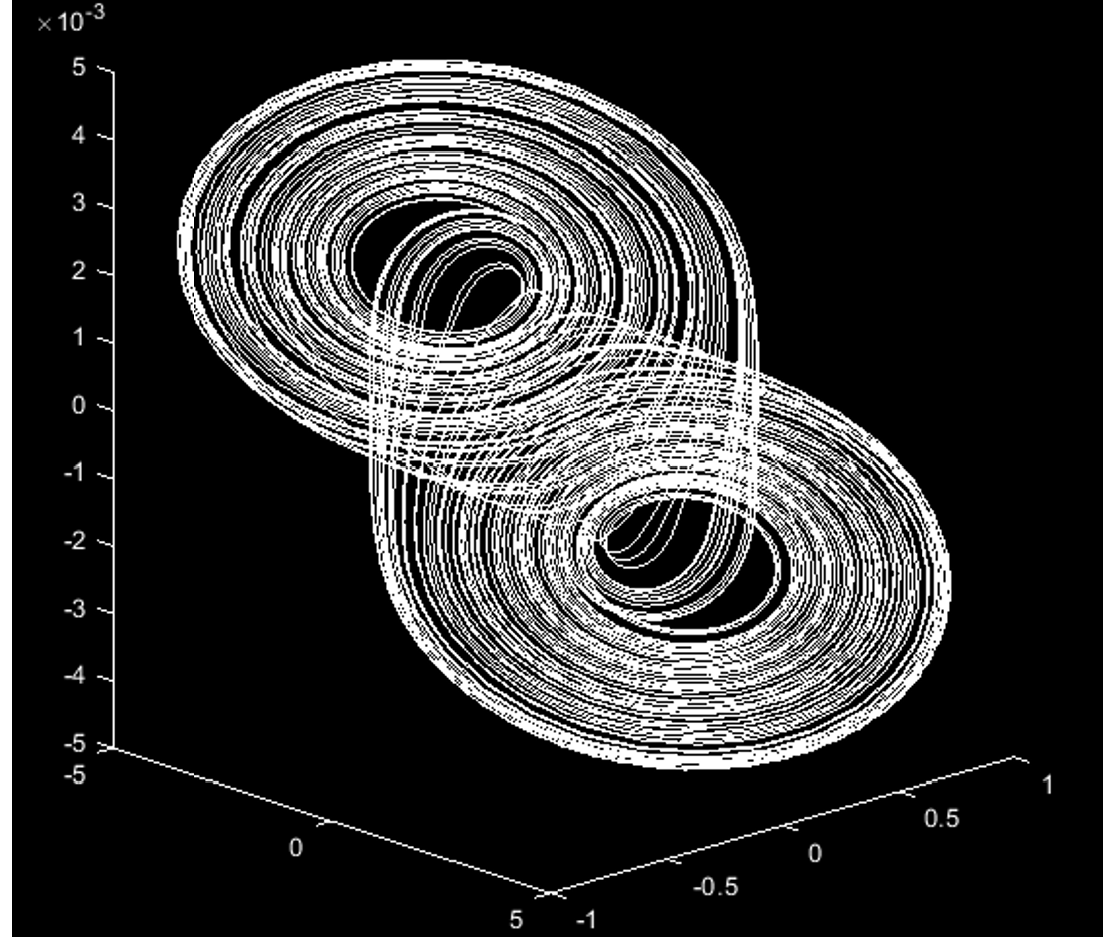}} 
 
    \caption{(a) Chua's circuit (n) MATLAB simulation results of attractor for Chua's circuit}
    \label{Chua}
\end{figure}

Chua’s circuit is a non-repeating chaotic system that requires a nonlinear element, a locally active resistor, and three or more energy storage elements \cite{fortuna2009chua}. In contrast to other common chaotic equations, like the Lorenz equations, Chua's circuit is more sensitive to its initial conditions. By varying the value of one of the parameters from 1.0 to 1.1, the trajectories of x(t) are extremely different. Most importantly, the Chua's circuit is "one of the simplest robust experimental proof of chaos and can be easily implemented in different ways" \cite{fortuna2009chua}. As demonstrated by Chua's Fig. \ref{Chua}, the butterfly effect and double scroll attractor are shown, portraying the chaotic behavior and three non-linear ordinary differential equations. Chua's circuit, with its three unstable equilibrium points, exhibits more chaos than prior systems.

 \begin{figure*}[ht]
    \centering{\includegraphics[width=5in]{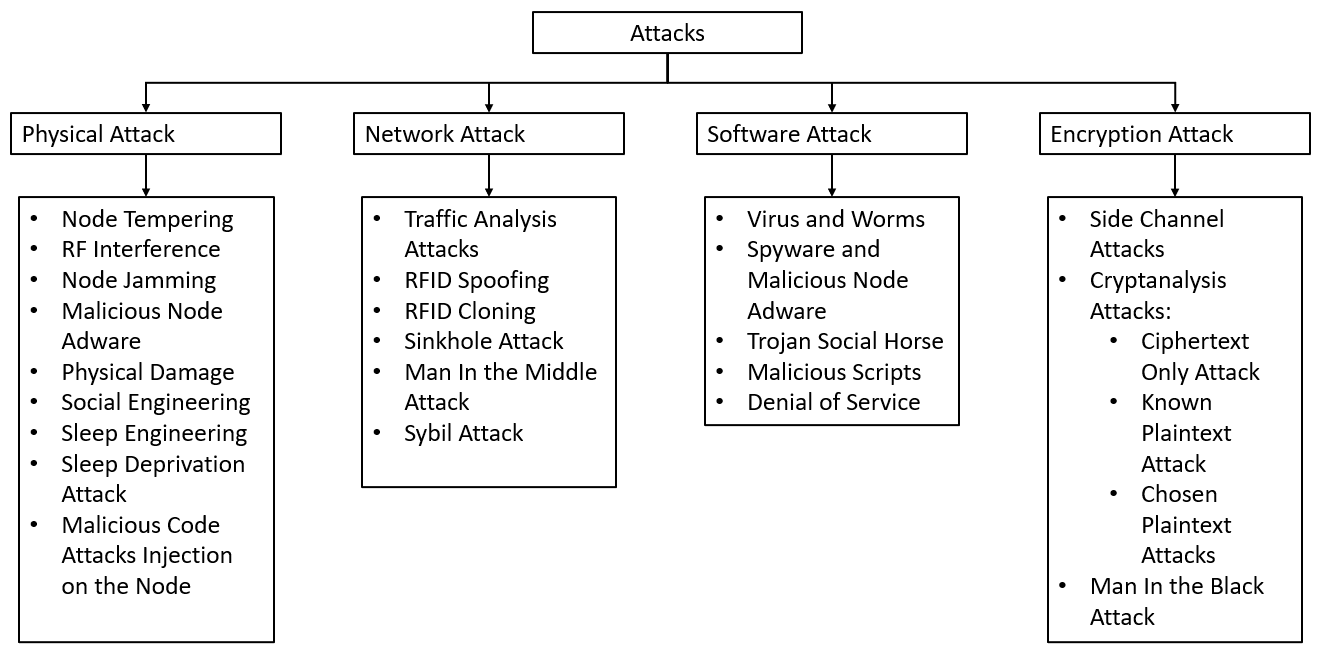}}
    \vspace{0.5 cm}
    \caption{ Security attacks \cite{8058363}
        \label{securityAttacks}}
\end{figure*}

With the traditional encryption algorithms, the IWDs achieve the designated tasks; however, they are not considered reliable when it comes to the security of the personal data generated by the device or its sensor, and they suffer so many security breaches. The process of making IWDs secure and reliable is complex, as they require security to be implemented at each layer, as Fig. \ref{securityAttacks} shows potential risks. Some of the security breach examples force the IWDs industry to advance research in the field of cryptography. These include a recent data leak that exposed the data of Fitbit and Apple’s over 61 million users from around the world \cite{dataleak}. For IWDs, vulnerable attacks are targeted on cardiac implants \cite{4531149, 10.1145/2991079.2991094}; implantable insulin delivery pumps \cite{6026732, radcliffe2011hacking}; and neurological implants \cite{PYCROFT2016454}. Thus, this encourages research toward improving the traditional encryption algorithms in terms of hardware utilization as well as making them a secure and robust encryption architecture.

In this paper, we will be focusing on physical attacks and encryption attacks. As our device is not dependent on RFID, network attacks will not be discussed in this paper. Also, due to the hardware encryption of the design, no data can be extracted from the encrypted message in the public channel. The attacker needs the exact copy of the circuit implemented to be able to decrypt the data. We will also not discuss the software attacks in which the attacker performs the attack by using viruses, worms, spyware, adware, etc. to steal data or deny the services which are not applicable to our hardware-oriented design. 

\section{Physical Attacks on Wireless Devices}
Physical attacks are concentrated on hardware devices in the system. 

\subsection{Node Tampering}
The adversary can perform the attack on the IoT system by damaging or tampering with some nodes. In this attack, the attacker physically alters the compromised node and can obtain sensitive information such as the encryption key. To tamper with nodes in our design, the attackers need access to the physical location of the transmitter and receiver circuits. They also need to extract the schematics of the transmitter and receiver to extract the node.

 \begin{figure}[!b]
     \centering 
\subfloat[]{\includegraphics[width=0.93\columnwidth]{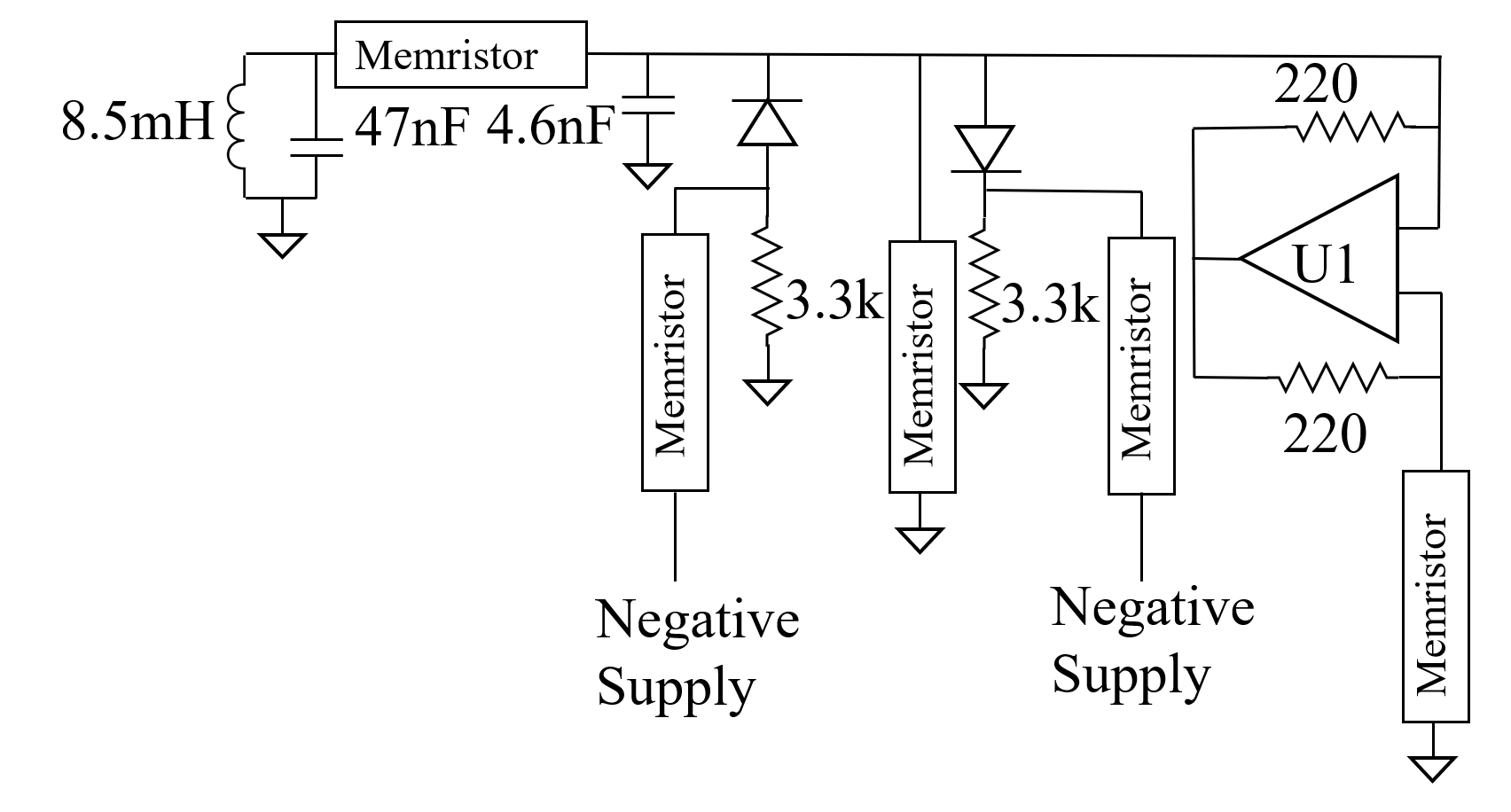}} \\ 
\subfloat[]{\includegraphics[width=0.93\columnwidth]{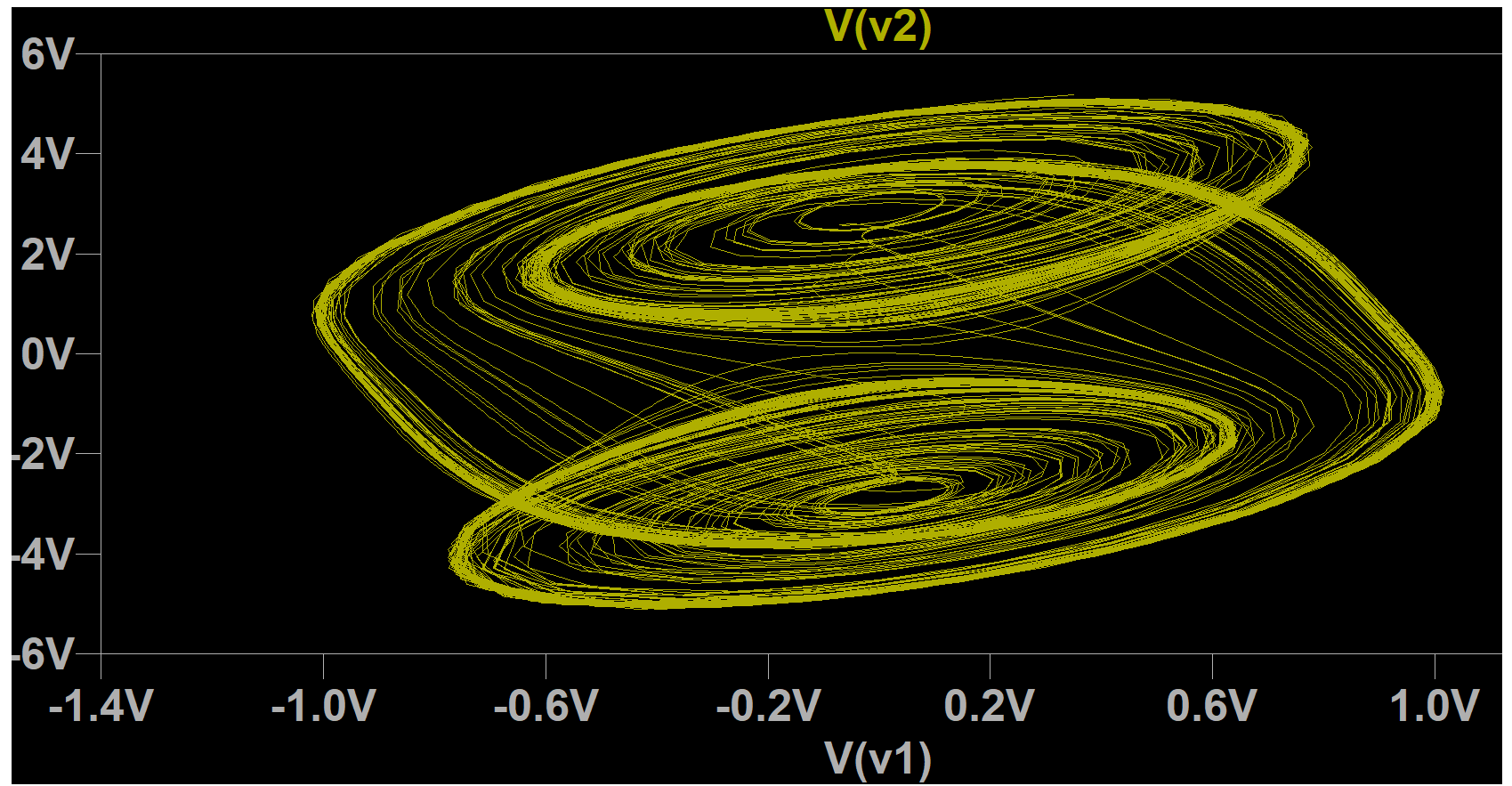}} 
    \caption{(a) Memristor-based Chua's circuit (b) Double scroll attractor for memristor U3 in Chua’s chaotic circuit  with memristor values $R_{on}$=0.8K $R_{off}$=1.65K $R_{init}$=1.65K D=70N uv=10F p=1
    \label{Chuamem}}
\end{figure}

Assuming access to the schematic and physical nodes in an untrusted foundry, we resolve this issue by replacing the resistors in the circuit with memristors. A memristor \cite{chua1971memristor} is a two-terminal electrical component relating electric charge to magnetic flux linkage. It can be viewed as a form of non-volatile memory that is based on resistance switching, which increases the flow of current in one direction and decreases the flow in the opposite direction. Here we use a logic locking mechanism \cite{rezaei2019hybrid, rezaei2020rescuing, rezaei2022DLE} based on the use of memristors as the keys for securing the circuit design against untrusted foundries. A Chua's circuit implemented with memristors is shown in Fig. \ref{Chuamem}a. The chaotic behavior of our designed circuit is shown in Fig. \ref{Chuamem}b as a double scroll attractor by plotting I-V characteristics for memristor U3 in the circuit. The values of the other memristors are tuned in a range such that they satisfy the chaotic equations and exhibit chaotic behavior.

\begin{figure}[!t]
    \centering{\includegraphics[width=3.4in]{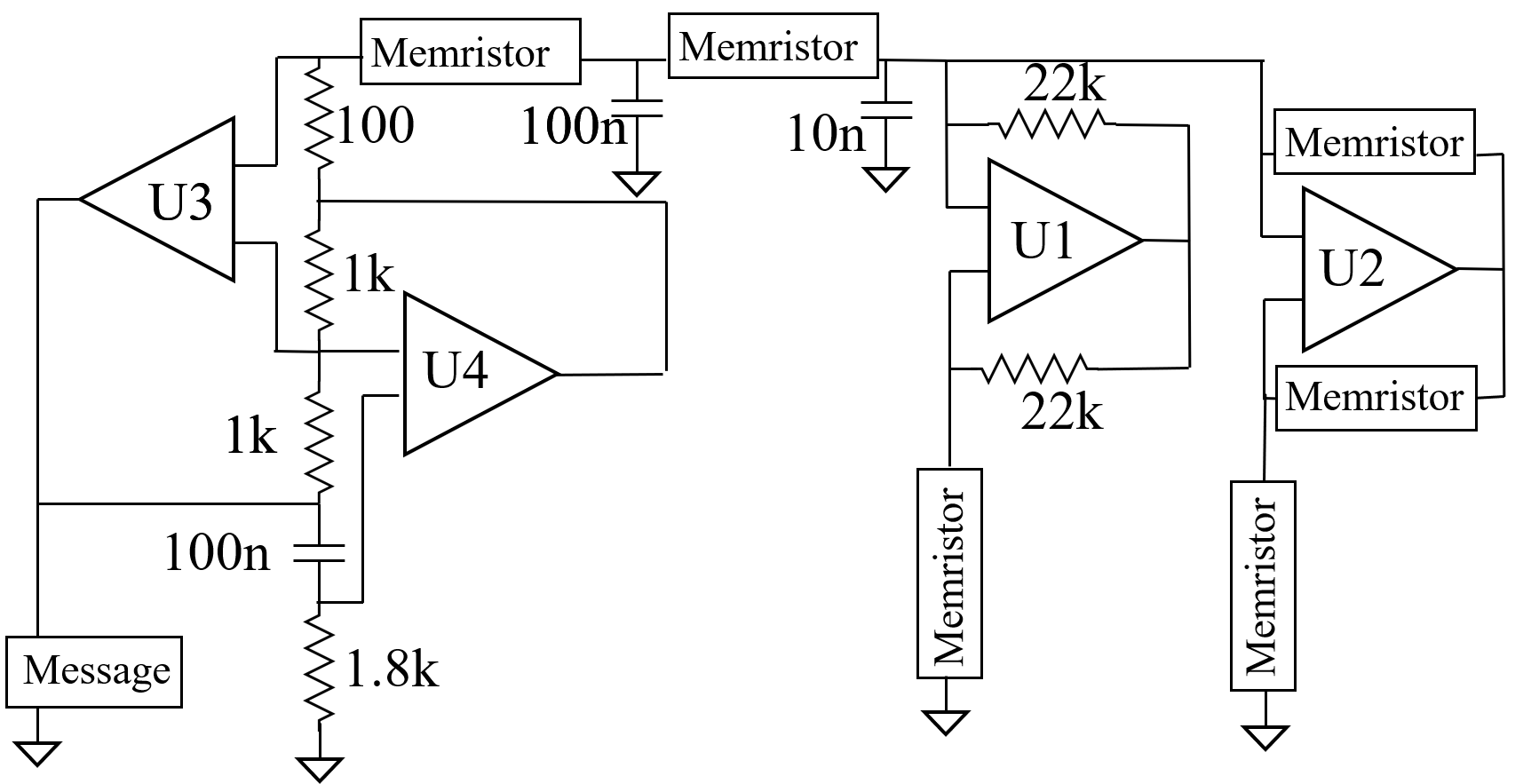}}
    \vspace{0.1 cm}
    \caption{The transmitter and receiver design of the communication system With this scheme, the message is encrypted by the sender, moved through a public channel, and decrypted by the receiver
        \label{TR}}
\end{figure}

\begin{figure}[!t]
    \centering{\includegraphics[width=0.95\columnwidth]{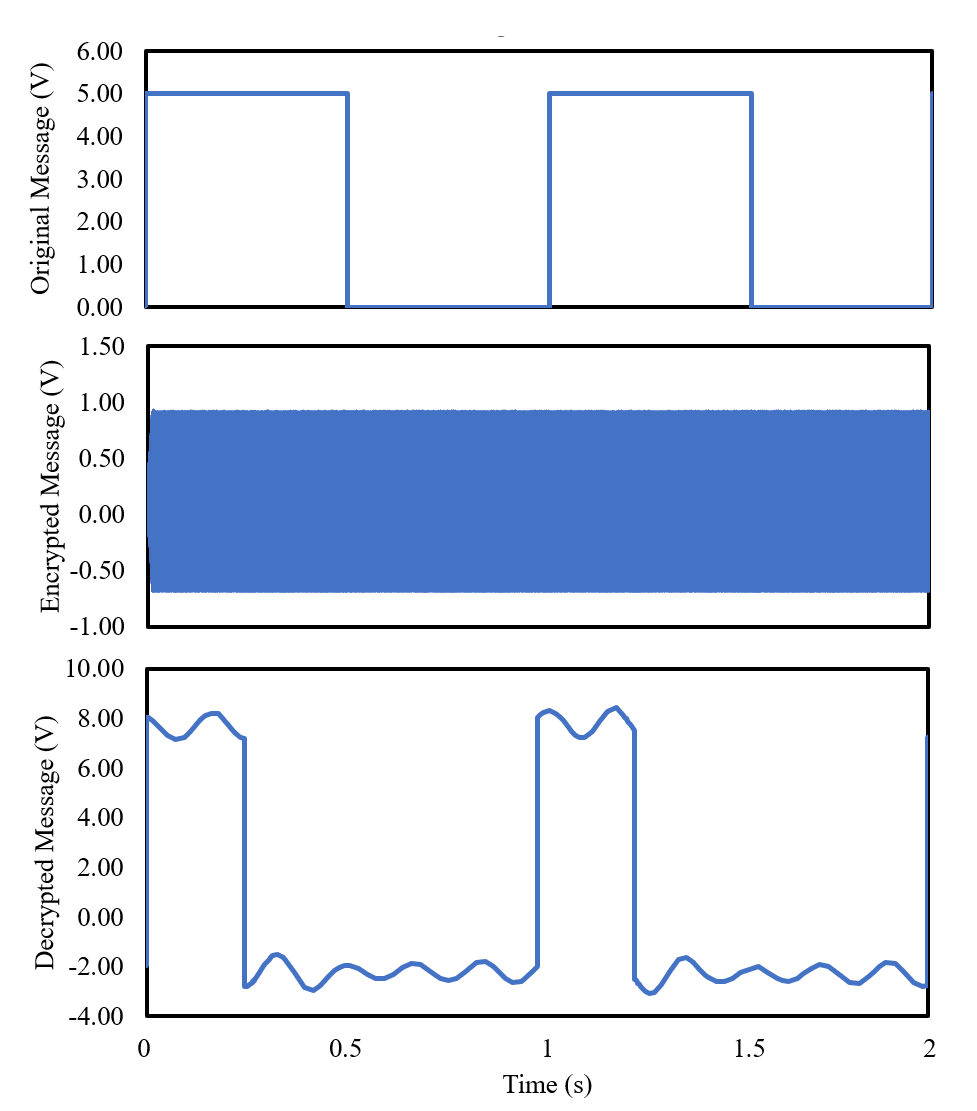}}
    \vspace{0.1 cm}
    \caption{Original, encrypted, and decrypted message for memristor-based Chua’s chaotic circuit
        \label{MED}}
\end{figure}

Most of the researchers have focused on designing a system to reliably transmit information from the sender to the receiver by making improvements to the way the chaotic circuit is designed. Here, to have a transmitter and receiver scheme, we replace the resistors in Chua's design with memristors, as shown in Fig.\ref{TR}. Fig. \ref{MED} shows the simulation of chaotic communication using Chua's memristor-based circuit.

By using memristors, a reliable and secure chaotic transceiver that is resistant both to eavesdroppers and untrusted factories is achieved. In this case, even if the attacker has the schematic of the design, stored from the foundry, they will not have the value that the memristor is set on, hence unable to replicate the design. The approach is effective due to the large key space created by implementing memristors in the circuit, along with the logic locking.

When working with memristors, however, process variations of these devices should be taken into account. Due to the non-deterministic nature of the Nanoimprint Lithography (NIL) used to form the electrodes and crossbars, a significant amount of process variability exists in memristors \cite{Chaudhuri}. The deposition and etching procedures involved in forming the thin oxide film \cite{etch}, and via-etching through the dielectric also make memristors particularly prone to variations. The ensuing misalignment, size mismatch, and change in dopant concentration inside the dielectric, breaks, openings, and shorts have a negative impact on the behavior of the memristor device and consequently, our security synchronization \cite{Pang}.

The chaotic behavior, however, is not just bound by the strict values of the memristors. Driving eigenvalues of the Jacobian matrix for chaotic equations as shown in \cite{Brown}, chaotic behavior is seen for a range of values instead of a single value. Hence, despite small changes in memristor value due to process variation, this will not affect the overall chaotic behavior of our system.

\subsection{RF Interference and Node Jamming}
\begin{figure}[!t]
    \centering{\includegraphics[width=0.95\columnwidth]{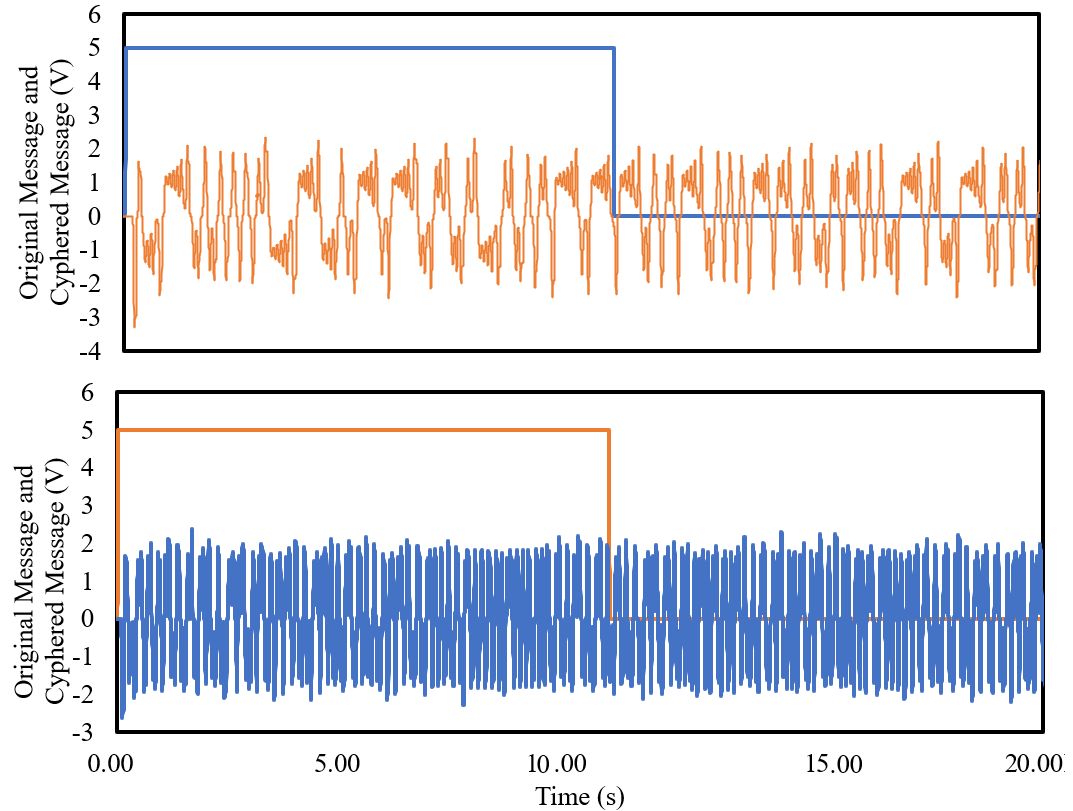}}
    \vspace{0.1 cm}
    \caption{Frequency change by changing coefficients in Chua's equation
        \label{RF}}
\end{figure}

The attacker can perform a Denial-of-Service (DoS) attack by sending noise signals over radio frequency signals, or by using a jammer, the attacker can disturb the wireless communication. The jammer emits a radio signal and can be implemented using either a waveform generator that continuously sends a radio signal \cite{Channel} or a normal wireless device that continuously sends out random bits to the channel. One method to address this issue is spatial retreats, in which jammed nodes try to evacuate jammed regions. Spatial retreats are suitable for mobile sensor networks. The spatial movement of our ciphered signal here is related to Chua's equation as follows:
\begin{align}
& x'=\sigma (y-x - f(x)) \nonumber\\
& y'=x - y + z \\
& z'=-\beta y \nonumber
\end{align}
To show that certain parameter choices lead to asymptotic stability rather than chaotic behavior, one can use a Lyapunov function as shown in \cite{daniel}. Solving these functions proves that we can achieve chaos with a variety of $\sigma$ and $\beta$ parameters. In our design, changing these parameters results in a change in the cyphered signal frequency. In Fig. \ref{RF}, the input is a 5-second pulse, and the high-frequency ciphered signal is also shown. The top picture shows a chaotic ciphered signal with a frequency of 20-100 times the frequency of the message. The $\sigma$ and $\beta$ parameters are then changed by adjusting the resistances in the circuit which increase the frequency of the chaotic ciphered signal to 80-400 times the frequency of the message as seen in the bottom picture Fig. \ref{RF}. The deciphering of both signals is done successfully in this case, showing that in the event of jamming or RF interference by the attacker, a dynamic frequency change can be performed to move the ciphered signal in the public channel to a safe region.

\subsection{Physical Damage} 
With the attacker having access to the physical locations of the transmitter and receiver, they can be physically damaged. However, since our transmitter and receiver are aimed to be used as ICs, they can be replaced for only a few dollars. When designing ICs the Non-Recurring Engineering (NRE) costs, such as teams, equipment resources, and materials other than the ICs, can add up to a couple of million dollars. However, ICs produced in the range of a million parts per wafer will divide the cost by the number of ICs produced per wafer. The costs of fabrication processes and NRE costs to build commercial ICs are therefore divided by millions; thus, the price of commercial ICs, produced in millions, does not go much higher than a few dollars. 

\subsection{Social Engineering} 
The attackers can physically interact with and manipulate users of an IoT system to obtain sensitive information to achieve their goals. This attack is not applicable to our design as the user is not aware of the schematic details of the transmitter and receiver. Unlike many software security systems, which rely on the user providing a password, in our hardware-based ciphering, the user does not have access to any sensitive information.

\section{Encryption Attacks on Wireless devices}
These attacks depend on destroying encryption techniques and obtaining the private key.

\subsection{Side Channel Attacks}
\begin{figure}[ht]
    \centering{\includegraphics[width=0.95\columnwidth]{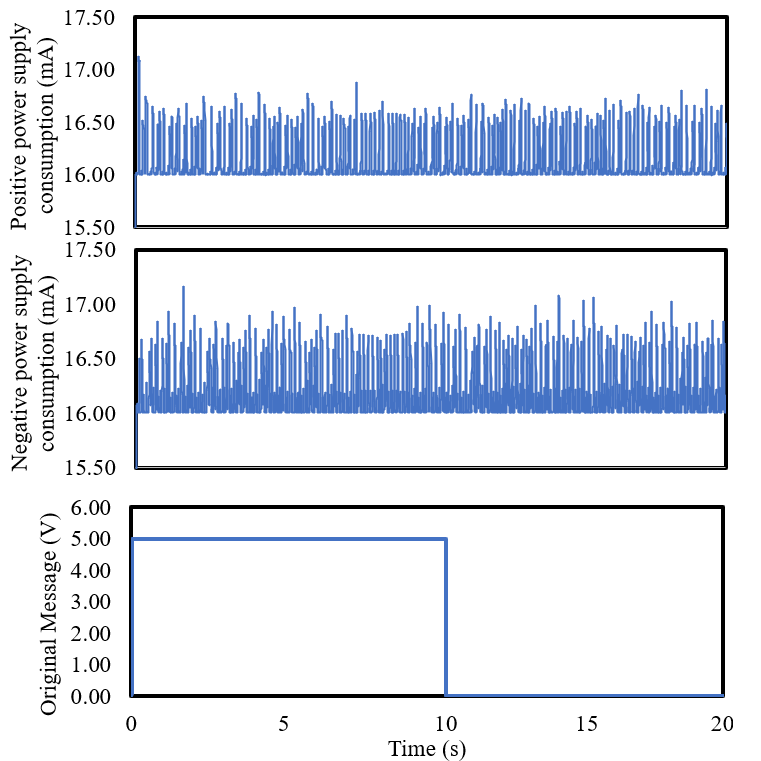}}
    \vspace{0.1 cm}
    \caption{Current consumption of the system shows no relevance compared to the message
        \label{Power}}
\end{figure}

The attacker takes advantage of the side-channel information emitted during the encryption process on devices. This information is distinct from both the plaintext and the ciphertext, and it can include details about the power consumption, operation time, fault frequencies, etc. With this information, the attacker can uncover the encryption key. In our design, one method for a side-channel attack is the attacker plugging into the power line and trying to extract the message pattern from the power line. Due to the isolation of the message from the power line, as shown in Fig. \ref{Power}, no correlation is seen between the positive and negative power lines and the message. As another scenario, the side-channel attack can be mapped as a Return Map (RM) attack, discussed in the next subsection, which is a method of monitoring the transmitted state's local minimum and maximum points to distinguish changing system characteristics.

\subsection{Ciphertext-Only Attack}
The RM is a type of Ciphertext-Only Attack (COA). RM attack can be defeated by performing the encryption modulation in a time-scaling factor. Chaotic Shift Keying (CSK) security vulnerabilities to RM attacks can be mitigated using a ``time scaling function", $ \lambda (x(t),m)$ to encrypt the plain-text message, $m(t)$. For any autonomous dynamical system,

\begin{equation} \label{eq8}
\frac{d}{dt}x= f(x)
\end{equation}

\noindent where the time scaling function is defined by,

\begin{align} \label{eq9}
\frac{dt}{d\tau}=\lambda(x) \nonumber\\
 \tau(t_{0})=\tau_{0}\\ 
 0<\lambda(x)<\infty\nonumber\\
\nonumber \end{align}

\noindent It is clear that then,

\begin{equation} \label{eq10}
\frac{d}{d\tau}x= \lambda(x)f(x)
\end{equation}

From equation (\ref{eq9}) $ \tau $, is strictly monotonic and increases with time, $t$ \cite{sampei1986time}. From equation (\ref{eq10}), the $ \lambda(x) $ does not alter the phase space of $x$ with respect to its attractors or equilibrium, and the only effect here is reaching a stable time \cite{materassi2008time}. The RM attack can be thwarted by encoding the message, $m(t)$, with the time scaling factor. This occurs due to the dependence on the variation of the state phase space.

We apply this time-scaling factor to a Chua's chaotic system by using a function, $ \lambda $(x,m), to create a Time Scaling Chaotic Shift Keying (TS-CSK) encryption system. Here, x is the transmitter parameter, and z are the receiver parameters. The system equations then become,

\begin{equation} \label{eq11}
\begin{array} {cc}
\dot{x}_{1}  = \sigma(x_{2}-x_{2}-f(x))\lambda(x,m) \\\dot{z}_{1}  = \sigma(z_{2}-z_{1}-f(x))\lambda(z,0)\\
\dot{x}_{2}  = (x_{1}-x_{2}+x_{3})\lambda(x,m) \\\dot{z}_{2}  = ((z_{1}-z_{2}+z_{3})\lambda(z,0) \\
\dot{x}_{3}  = (-\beta x_{2})\lambda(x,m) \\\dot{z}_{3}  = (-\beta z_{2})\lambda(z,0)
\end{array}
\end{equation}

\noindent where,

\begin{equation} \label{eq12}
\lambda(x,m) = 
\left\lbrace \begin{array} {cc} \lambda _{m} & if \hspace{0.5cm} \delta_{x}=0 \\ \lambda _{1-m} & if \hspace{0.5cm} \delta_{x}=1  \end{array} \right\rbrace
\end{equation}

\noindent Here $ \delta (x)$ is the decision engine. To have stable security, $ \delta (x)$ should be selected such that the switching event of $ \lambda (x,m)$ cannot be decoded or extracted from the signal that is being communicated. There are several options for $(x)$ that satisfy these conditions. \cite{materassi2008time}, uses $ \delta (x)$ such that,

\begin{equation} \label{eq13}
\delta(x) = 
\left\lbrace \begin{array} {cc} 0 & \frac{\upsilon^{T}x}{h} is even \vspace*{6pt}\\ 1 & \frac{\upsilon^{T}x}{h} is odd  \end{array}\right\rbrace
\end{equation}

\noindent where the unitary selection vector $ \upsilon $ is,

\begin{equation} \label{eq14}
\Phi _{sync} =
\left[\begin{array}{cc} \upsilon _{1} \\ \upsilon _{2} \\  \upsilon _{3} \end{array}\right] 
\end{equation}

However, the implementation of this decision engine gets increasingly difficult and expensive when the oscillation rate of Chua's system is higher. Based on these considerations and starting with the idea of simplifying to two regions, the initial decision engine was designed. 

\begin{equation} \label{eq15}
\delta(x) = 
\left\lbrace \begin{array} {cc} 0 & \upsilon^{T}x<0 \\ 1 &  \upsilon^{T}x\geq0  \end{array} \right\rbrace
\end{equation}

The decision engine described by equation \ref{eq13} is immune to return map attacks \cite{materassi2008time}. As the return map does not modify the underlying Chua's function in correlation with the orbital foci, TS-CSK encryption mitigates the RM attack. As long as $x$, for the TS-CSK system of equation (\ref{eq11}), is chosen so that the underlying Chua's system is chaotic, then the system is protected from RM attacks. Return time-map attacks require the message, $m(t)$ to somewhat obscure the time difference between orbits. It must be noted that this immunity does not hold for  plus-minus TS-CSK (PM TS-CSK) decision engines with a reasonable time difference between bits, instead requiring a marginally more complicated decision engine. This is borne out the system equations,

\begin{equation} \label{eq16}
\delta(x) = 
\left\lbrace \begin{array} {cc} \delta_{z} & x_{2}(t)<-\sqrt{\rho(\beta-1)} \vspace*{3pt}\\ 1-\delta_{z} & -\sqrt{\rho(\beta-1)}\leq x_{2}(t)< 0 \vspace*{3pt}\\ \delta_{z} & 0<x_{2}(t)< \sqrt{\rho(\beta-1)}\vspace*{3pt}\\ 1-\delta_{z}  &  x_{2}(t)\geq\sqrt{\rho(\beta-1)} \end{array}\right\rbrace 
\end{equation}

\noindent where,

\begin{equation} \label{eq17}
\delta_{z} = 
\left\lbrace \begin{array} {cc} 1 & x_{3}(t)\geq\beta-1 \\ 0 &  x_{3}(t)<\beta-1  \end{array} \right\rbrace
\end{equation}

The regions of operation and switching about the equilibrium of the system are derived from this new decision engine. When the third dynamic, $ x_{3} (t)$, crosses from one side of its equilibrium to  the other the system switches. The system splits the second dynamic into four regions to increase the number of regions of operation. These regions are below the negative focus, between the negative focus and the origin, between the origin and the positive focus, and above the positive focus. Equations (\ref{eq16}) and (\ref{eq17}) describe the 8-section TS-CSK system.

\subsection{Known-Plaintext and Chosen-Plaintext Attacks}
These attacks are derived from inserting a known plain-text message and comparing it to the responding ``ciphertext". Any circuitry-implemented CSK system is immune to this type of attack, as the message will just extract the orbitals or orbital speed system for the CSK and TS-CSK frameworks responsively. This attack, which is a repetition of the ciphertext, is only useful when the same initial conditions of the system with the transmitter can be guaranteed, which is impossible within a circuit without internal knowledge of Chua's system and significantly accurate measurements of the state values.

\section{Conclusion}
Resource-limited devices depend on low-power modes of security to transmit data. Due to the dependence of synchronous and asynchronous encryption on power-consuming processors, implementing these modes of security on resource-limited devices is becoming more challenging day by day. To develop a new mode of security, chaotic encryption is gaining more interest. This work contributes toward the goal of achieving the efficient chaos ciphering implemented on the chip along with the sensors to encode the data at its very origin. 

The study of attacks on hardware-implemented Chua's chaotic equations with simulation in multiple platforms, i.e., MATLAB Simulink and LTSpice enables a closer move 
to the desired security level that can 
operate in a post-quantum era. 
In this work, we thoroughly examined the attacks, in the two subgroups of physical and cryptographic attacks, on continuous chaotic systems in devices with limited resources, as well as any viable defenses through simulations. A promising method simulated in this paper is the use of memristors to avoid node tampering. Memristors are capable of acting as the keys for securing the circuit design with negligible power consumption added to the circuit. The paper also proves in simulation the robustness of Chua's encryption to common side-channel attacks. 

\section*{Acknowledgment}
This 
research is based upon work supported by the National Science Foundation under Grant No. 2131156.

\end{document}